# Recommending Influenceable Targets based on Influence Propagation through Activity Behaviors in Online Social Media

Dhrubasish Sarkar[(✉)]

Amity Institute of Information Technology (AIIT), Amity University, Kolkata, India

dhrubasish@inbox.com[2]

## ABSTRACT

Online Social Media (OSM) is a platform through which the users present themselves to the connected world by means of messaging, posting, reacting, tagging, and sharing on different contents with also other social activities. Nowadays, it has vast impact on various aspects of industry, business and society along with on user's life. In an OSN platform, reaching the target users is one of the primary focus for most of the businesses and other organizations. Identification and recommendation of influenceable targets helps to capture the appropriate audience efficiently and effectively. In this paper, an effective model has been discussed in egocentric OSN by incorporating an efficient influence measured Recommendation System in order to generate a list of top most influenceable target users among all connected network members for any specific social network user. Firstly the list of interacted network members has been updated based on all activities. On which the interacted network members with most similar activities have been recommended based on the specific influence category with sentiment type. After that the top most influenceable network members in basis of the required amount among those updated list of interacted network members have been identified with proper ranking by analyzing the similarity and frequency of their activity contents with respect to the activity contents of the main user. Through these two continuous stages an effective list of top influenceable targets of the main user has been distinguished from the egocentric view of any social network.

## 1.0 Introduction

Since last few years the Online Social Media (OSM) have drawn huge attention of its users and grown enormously. Nowadays, millions of users across the world consider these platforms as a part of their daily lives spending a huge amount of their time to create and observe many information. Through these platforms the users can create or exchange their ideas or thought through textual, visual and web contents as well as they can react, comment, share and tag others on those contents of own and other users. Hence these Online Social Media (OSM) platforms have drawn the focus of the data driven business industries to understand or gain some valuable information about the customers, their preferences, choices and various influencing parameters etc. by analyzing their social activity contents, it's types and frequency.

There are generally two types for social network observation, such as egocentric and sociocentric [1]. In sociocentric approach the complete or whole network is considered but egocentric approach considers on individual (called ego) network with focusing the interactions among that individual person with all connected people (called alters) indirectly or directly. In our current paper, a effective algorithmic mathematical technique has been discussed by incorporating an efficient Influence Measurement [2, 3] with a significant recommendation system in order to generate a list of top most influenceable target users among all interacted network members for any specific social network user (main user) in egocentric online social network.

At first a list of interacted network members are selected based on their acitivities with the main user and then all those interacted network members are categorized based on the activity contents classification (five classes) and sentiment (two classes) analysis on them for all five types of social activities. After that the list of interacted network members (based on activities) again has been updated with a less value by the operation of a significant recommendation system of textual lexical analysis (similarity and frequency), on which the interacted network members with most similar activities have been recommended with proper ranking by analyzing the recommendation of their post contents with respect to the post contents of the target user. Then the top most influenceable network members among those updated list of interacted network members have been identified in any specific influence category with specific sentimant type based on the required amount.

The proposed model can work on three most popular social network platforms of the main user separately; Facebook[1], Twitter[2], LinkedIn[3] and only considers the egocentric scenario as it is useful when the research focuses on individuals in the network compared to the complete network analysis. The chapter is organized with a section in next discussed about the previous works done in this area as Literature Review. After that the following sections are covered with the Data Collection, Preprocessing, Influence and Activities Analysis, and the proposed Recommendation System. Last two sections contain the conclusion and future scope.

## 2.0 Literature Review

Network structure, its properties and analysis are also very much common in the various fields of business and industry. An OSN structure can be described as a graph or network structure which comprises the collection of individuals and the links or association between them. It is very much essential to identify the most important targets in a networked environment in order to address various issues and services related to the network. Primarily the graph theory concepts are used for topological or structural analysis of social networks. Social activities analysis of the connected users also help to explore the network with insight knowledge.

Influenceable targets identification in OSNs has huge significance as far as its business applications are concerned. It can help a business or organization to reach to its target audiences. Hence this domain draws the attention of many researchers. Identifying influenceable targets is basically influence maximization problem. Many mathematical and algorithmic techniques have been proposed to identify influenceable targets or users in a network.

Domingous and Richardson [4] did the first study in this research field. In their study, instead of viewing a market as a set of independent entities, they considered it as a social network and represented it as Markov random field in their work. They developed three algorithms for determining influential users which are be used in viral ways of marketing. They have also discussed the reviews on some of the significant approaches used to identify influential targets have been presented. Kempe et al. [5] formulated influence maximization as a discrete optimization problem and proved that this optimization problem is NP-hard. Through experiments they showed that the proposed algorithm could significantly outperform the classic degree and centrality based heuristics in influence maximization. Bonchi [6] took his work to show the diffusion of influence using a data mining perspective. However, Saito et al. [7, 8] revealed that the diffusion of the different topics is not same because of the different preferences of the users that will affect their role in the spread of a specific topic. Later, Tang et al. [9] proposed the topic-level social influence using Topical Affinity Propagation (TAP) for large social networks. Later, Zhou et al. [10] proposed top-K influential nodes mining on the basis of user choices using a two-stage mining algorithm. Li et al. [11] proposed to consider the individual behaviors of persons to model the influence propagation in heterogeneous social networks. Marsden [12] discussed about centrality measures on egocentric and sociocentric networks. Later, Chung et al. [1] mentioned about egocentric and sociocentric approaches on social network analysis with highlighting the scopes or challenges for both cases.

Previously similar recommendation model already proposed by Debnath et al. [22] to recommend trending authors considering their research activities in google scholar. To measure users' activities in OSNs, Debnath et al. [13] analyzed the lexicons and citation parameters for particularly Twitter network users' activities. Debnath et al. In egocentric OSNs, they proposed for identification of top-k number of influencers considering only the activity behaviors [2]. In another [3], they considered the network structure of the user with it's network members including analysis of activity behaviors among them and proposed a model based on egocentric OSNs in order to recognize influential users as nodes for that user.

## 3.0 Dataset Collection Process with Details

As the input for our algorithmic method massive amount social media contents of a long time duration were needed for any particular user with details from different social media platforms like *Facebook, Twitter, LinkedIn* etc. So here from past two years all social media activities data have been collected for a particular user with it's network members associated by activities and prepared nine datasets of different contents for each social network separately. The sequential process is mentioned below (shown in Fig. 1).

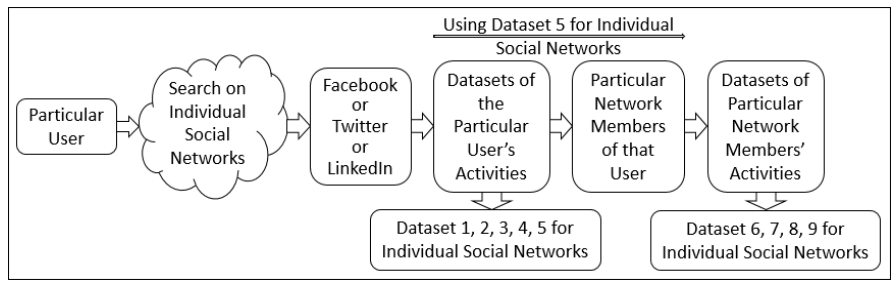

**Fig. 1.** Diagram of Dataset Collection Process

## 3.1 Main User's Activities Data

For individual Online Social Media (OSM) platforms, from the activity log portal all activity contents with details have been collected within a long time duration for the main user. Datasets are such as; all contents of own posts, commented posts, reacted posts, tagged posts, shared posts (*Dataset 1*), all additional contents of shared posts (*Dataset 2*), all comments of commented posts (*Dataset 3*), all messages of the conversations with other network members (*Dataset 4*), the list of all the network members associated with all main user's activities like posting, reacting, commenting, sharing, tagging, messaging activities (*Dataset 5*).

## 3.2 Network Member's Activities Data

Here also for individual Online Social Media (OSM) platforms, from the activity log portal all activity contents with details have been collected within a long time duration for all network members of that main user. Datasets are such as; all contents of own posts, commented posts, reacted posts, tagged posts, shared posts (*Dataset 6*), all additional contents of shared posts (*Dataset 7*), all comments of commented posts (*Dataset 8*), all messages of the conversations with their other network members (*Dataset 9*).

## 3.3 Tools and Libraries for Data Collection

For this '*Data Collection*' stage, as mentioned by Fredrik and Stolpe [14] different libraries for data crawling from social media are useful. So from individual social media platforms (*Twitter, Facebook, LinkedIn*), all datasets (excluding *Dataset 3* and *8*) can be crawled using Twitter4J'[4] by Yusuke [15], 'Facebook4J'[5], 'Graph API'[6], 'LinkedIn4J'[7], 'REST API'[8] respectively. But these libraries (specially all 4Js) are not able to get comment/reply thread directly from any particular post, therefore, for *Dataset 3* and *8* the '*Web Scraping*' method using 'jsoup'[9], a java library by Hedley [16] is forever applicable.

## 3.4 Details of the Datasets

Here firstly all post contents of individual user (main user and each interacted network member from dataset 5) are stored in different folders and also all network members' data are specially separated from target user's data. After that again, for all post contents of each user have been distinguished and stored in different csv files based on the different activity types of the particular contents, such as all post contents of four different activities (React, Comment, Tag and Share) have been stored in four different csv files inside that particular user folder for each user.

In these datasets, there are total eight colums of different attributs, they are; Post/Tweet ID, Post/Tweet Content, User Name, User ID, React/Favorite Count, Share/Retweet Count, Language, Time and also comments thread. A sample twitter data in text format mentioned below (shown in Fig. 2).

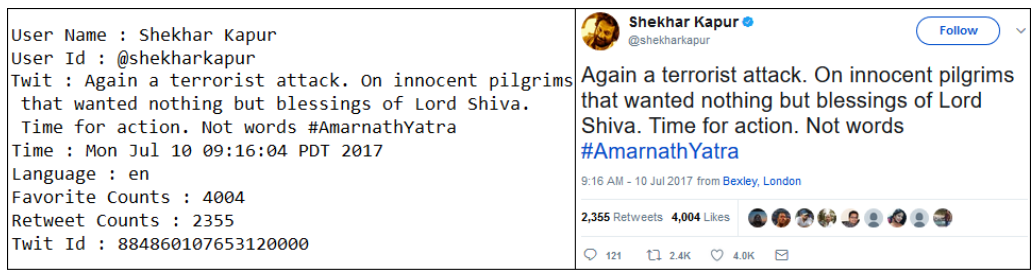

**Fig. 2.** Snapshot of a tweet data crawled from Twitter

## 4.0 Primary Preprocessing of Data

As we all know proper preprocessing works for making input data suitable based on algorithms produce better result. Therefore, four different textual preprocessing works on all datasets excluding '*Dataset 5*' have been incorporated here which are briefly discussed below (shown in Fig. 3).

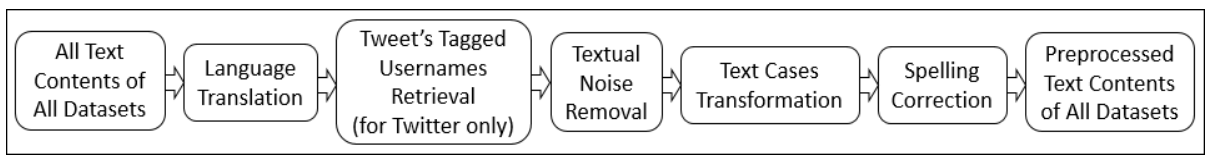

**Fig. 3.** Diagram of Data Preprocessing Process

4.1 *Language Detection and Translation:* Using 'TextBlob'[10] python library by Loria et al. [17], for all the social contents languages have been detected, if there exist non-English languages then they have been translated into English language.

4.2 *Tagged Tweeters Collection:* For tweets only, all words followed by '@' symbol represents username of tagged users. So all those words have been removed from tweets and tagged users have been added in '*Dataset 5*'.

4.3 *Textual Noise Removal:* All special symbols, set of special symbols excluding the single comma and full stop, emoticons which represents emojis, unrecognized characters for the attached images and GIF files have been removed from all social contents.

4.4 *Textual Spelling and Correction:* Using 'TextBlob'[10] python library by Loria et al. [17], all text cases have been converted into lowercase and all the noisy spelled words have been corrected for all social media contents.

## 5.0 Influence And Social Activities Analysis

Here methodologically, the influence has been tried to evaluate by analyzing the social activities of the main user and it's interacted network members from *Dataset 5* in a specific time duration. The sequential detailed process is mentioned below (shown in Fig. 4).

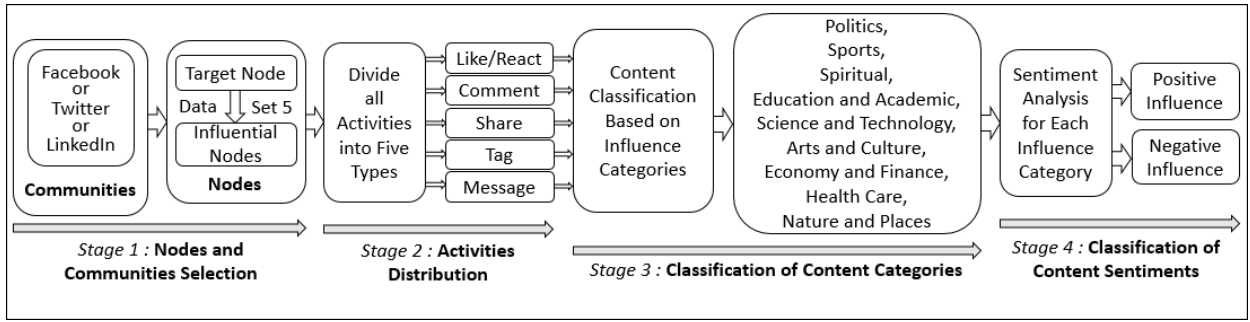

**Fig. 4.** Diagram of Influence and Activities Analysis Process

## 5.1 Step 1: Targets Selection from OSMs

Here firstly the list of all the interacted network members from *Dataset 5* have been only considered from where only the *influenceable targets* can belong because others are not associated with the *main user* throgh activities during that particular time period. This also reduced the complexity of our algorithm.

This *influenceable targets* and *OSMs* are like; friends, followers, connections for Facebook, Twitter, LinkedIn respectively with also following pages for all three OSMs. One notable point, the public and private groups can never be *influenceable targets* due to the privacy with approval feature by other known/unknow OSM user (as *admin*). The representation is mentioned below (shown in Table. 1).

**Table. 1.** Representation of Influenceable Targets and Online Social Media (OSM) platforms

| OSMs | Facebook | Twitter | LinkedIn |
|---|---|---|---|
| **Influenceable Targets** | Friends and Followers | Following profiles | Connections and Followers |
| | Facebook Pages | Twitter Pages | LinkedIn Pages |

## 5.2 Step 3: Categories Classification of Social Contents

Here all the social activities data from *Dataset 1* and *4* (main user's data) and *Dataset 6* and *9* (interacted network members' data) have been classified into five classes (*Technology, Politics, Sports, Business* and *Entertainment*) by the multi-class text classification with LSTM[11], a commonly used Recurrent Neural Network (RNN) using TensorFlow 2.0 framework.

This classification stage has not been applied to other dependent data from *Dataset 2* and *3* (main user's data) and *Dataset 7* and *8* (interacted network members' data) which are not owned by the post owners and dependent on *Dataset 1* and *6* respectively.

## 5.4 Step 4: Sentiments Analysis of Social Contents

Here all the social activities data from *Dataset 1, 2, 3, 4* (main user's data) and *Dataset 6, 7, 8, 9* (interacted network members' data) have been analyzed based on two sentiment classes (*Positive* and *Negative*) for each of the previous mentioned categories using VADER[12] (Valence Aware Dictionary and sEntiment Reasoner), a lexicon and rule-based sentiment analysis which uses the combination of sentiment lexicon and list of lexical features (example, words), labelled according to their semantic orientation; either positive or negative.

## 6.0 Recommendation System

In the proposed work firstly, the properties of all social activity contents of both the main user and it's interacted network members (**Dataset 5**) have been analyzed using the textual lexical analysis (similarity and frequency) on those activity contents.

Here the main aim was to recommend the highest similar post contents of the interacted network members with respect to all the post contents of the main user. After that a similar but short updated list of interacted network members (updated **Dataset 5**) with very similar activity contents with the main user has been derived. The detailed sequential stages are mentioned below (shown in Fig. 5).

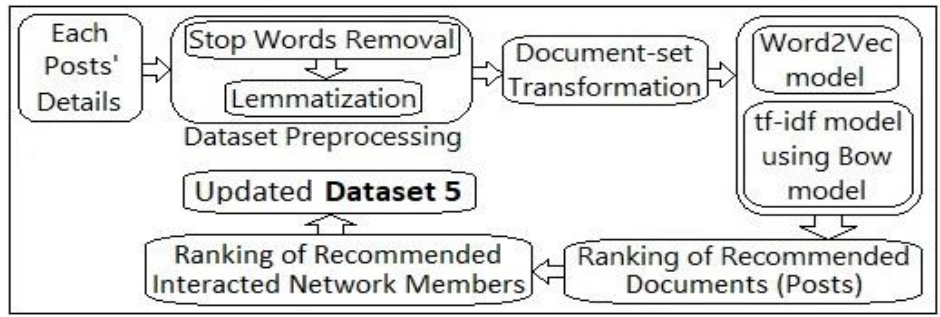

**Fig. 5.** Diagram of Recommendation Process

**6.1 Secondary Preprocessing of Data**

Here three different textual preprocessing works on all datasets excluding '*Dataset 5*' have been implemented here using 'Natural Language Toolkit (NLTK)'[13], a natural language processing package of python for only this recommendation algorithmic process. This stages are mainly responsible for removal of all unnecessary irrelevant details to get better recommended results for content analysis.

**Step 1: Stop Word Removal**

This stage is responsible for removal of all frequently occurred irrelevant words (*Stop Words*) from each post content which affect the recommendation indexing of post contents badly. Here all the post contents have been tokenized and all such words (example, *wh words, be and have verb forms, articles, prepositions, conjunctions* etc) using 'nltk.tokenize' and 'nltk.corpus' python packages respectively.

**Step 2: Lemmatization without Stemming**

This stage is responsible for transformation of all words retrieved from the previous stage into the root words for each post content which have reduced the word count to produce efficient recommendations. Therefore, we have applied *Lemmatization* due to it's morphological analysis process on words instead of *Stemming* as it sometimes generates nonexistent or unexpected words, but lemmatization always generates actual root words. Here 'WordNetLemmatizer' has been used which has built-in morphological python function of 'WordNet'[14] in the 'wordnet' module at 'nltk.stem' package.

**Step 3: Document-Set Transformation**

After completion of previous stages, actually a new optimized dataset has been generated as *Document-Set* where each post content of this data will be considered as an individual document. This individual post content (*document*) of all users will be considered as the key input data for further recommendation process.

## 6.2 Recommendation Analyzing Contents of Social Activities

The recommendation process has been proposed by forwarding four sequential tasks, followed on the data of the previously preprocessed dataset (*Document-Set*) with each individual post content (*document*) analysis and made more suitable outcomes/results for the proposed recommendation system.

**Step 1: Word2Vec Model Utilization**

This stage is responsible for generation of *word vectors* for all words of each post content/document from the Document-set mentioned previously using Google's pre-trained 'Word2Vec'[15] model and retrieval of *cosine similarity* values from the *word vector pairs* considering the *pairwise distances* among them. As this is one of the crucial stages for the betterment of our recommended results, therefore, the implementation of this model has been applied using 'gensim'[16] a python package popular for the operations in Information Retrieval (IR) and Natural Language Processing (NLP) domains.

**Step 2: BoW Model Utilization**

This stage is responsible for generation of a table structure which contains the *frequencies* or *number of occurrence* for all words in individual post content/document from the Document-set mentioned previously. Here for each post content/document from Document-se, it has been tokenized and then a dictionary has been created containing the frequencies of all words of that document. This is an implementation of *bag-of-words* (BoW) model where the rows represent documents and the columns represent frequencies of unique words.

**Step 3: Tf-idf Model Utilization**

This stage is responsible for generation of *vector values* for all words of each post content/document from the Document-set mentioned previously based on *tf-idf* (*term frequency-inverse document frequency*) model using 'feature_extraction' module of 'scikit-learn'[17], a python package popular for the operations in Machine Learning (ML). Here the first word 'term frequency' represents the same result like *bag-of-words* (BoW) model which is mentioned above and the second term 'inverse document frequency' represents the amount of non-occurrence for any word in any document.

**Step 4: Ranking of Recommended Users based on Documents**

Here, the output results from previous stages have been used to create a new updated list of interacted network members of the *main user* with influenceable ranking by recommending documents (post contents) based on similarity and frequency. The proposed recommendation algorithm (Algorithm 1) is explained below.

**Algorithm 1:**

**Let:** The total number of,

    Words in the Main user's documents (key post content) = m.

    Target users' documents (post contents) in the Document-set = N and Words in any of those documents (post contents) = n.

So, Key-term Word count (w) = 1 to m, Document count (d) = 1 to N, and Paper Word count (w) = 1 to n.

**Let:** For both the Main user's (mu) document and Target users' (tu) documents respectively,

    The vector values of any word from Word2Vec model are $Vw_{(mu)}$ and $Vw_{(tu)}$.

    The frequency values of any word from the table of BoW model are $BWw_{(mu)}$ and $BWw_{(tu)}$.

    The combined values of any word from tf-idf model are $TIw_{(mu)}$ and $TIw_{(tu)}$.

    The length of tf-idf values of any document are $TIl_{(mu)}$ and $TIl_{(tu)}$.

**Let:** The values between the Main user's document and Target users' document,

    The pairwise distance of two word vectors = $Vp_d$.

    The cosine similarity of tf-idf = $TI_{cs}$.

**Let:** The Normalized value and Recommendation Index value of any document are $N_d$ and $R^+_d$ respectively.

    *{Main user's document Evaluation}*

1:    Select each word of the Main user's document (key-term).
2:    for w = 1 to m do
3:        Evaluate, $Vw_{(mu)}$, $BWw_{(mu)}$ and $TIw_{(mu)}$.
4:    end for
5:    Repeat the same process until all done.
6:    Evaluate, $TIl_{(mu)}$ using all values of $TIw_{(mu)}$.

    *{All Post Contents Evaluation}*

7:    Select each of the Main user's documents (post contents).
8:    for d = 1 to N do
9:        Select each word of that document.
10:        for w = 1 to n do
11:            Evaluate, $Vw_{(tu)}$, $BWw_{(tu)}$ and $TIw_{(tu)}$.
12:        end for
13:        Repeat the same process until all done.
14:        Evaluate, $TIl_{(tu)}$ using all values of $TIw_{(tu)}$.
15:    end for
16:    Repeat the same process until all done.

    *{Rank Recommended Post Contents on Main user's document}*

17:    Select each of the Target users' documents (post contents).
18:    for d = 1 to N do
19:        Comparing each word of the Main user's document and that document.
20:        for w = 1 to m with w = 1 to n do
21:            Evaluate, $Vp_d$ using $Vw_{(mu)}$ and $Vw_{(tu)}$.
22:            Calculate, $N_d = (\ BWw_{(mu)} + BWw_{(tu)}\ ) / Vp_d$
23:        end for
24:        Evaluate, $TI_{cs}$ using $TIl_{(mu)}$ and $TIl_{(tu)}$.
25:    Calculate, $R^+_d = (\Sigma N_d \_ TI_{cs})$
26:    end for
27:    Repeat this previous procedure for all other documents.
28:    Sort the $R^+_d$ values of all the documents for all interacted network members in decreasing order.

## 7.0 Top Most Influenceable Targets Evaluation

Now here, all normalized recommendation indices ($R^+_d$) can be distinguished with both category and sentiment classes of influences separately for all Influenceable Target users with respect to the Main user and evaluated top most influenceable targets based on required amount by another algorithmic way (Algorithm 2) mentioned below where we eliminated some of the common influenceable targets having high default influence.

> **Algorithm 2:**
>     ***Step 1:*** Let assume the required amount is $N_{it}$ (50≤ $N_{it}$ ≤Network Size), then collect first $N_{it}$ number of *Influenceable Targets* from sorted list of Algorithm 1, as '*Selected Influenceable Targets*' based on their normalized Recommendation Indeces ($R^+_d$) on any categorical and sentiment classes of influence.
>     ***Step 2:*** Let the number of users with more than 5000 connections is $D_{it}$. They are removed considering as '*Default Influenceable Targets*' (example, verified pages/profiles) for already having huge influenceable impact in default.
>     ***Step 3:*** **Consider**, the remaining **($N_{it}$ - $D_{it}$)** number of users as '*Effective Influenceable Targets*' and replace the value of $N_{it}$ as ($N_{it}$ - $D_{it}$) by considering those as '*Top Most Influenceable Targets*'.

## 8.0 Conclusion

In this chapter, considering the Egocentric Online Social Networks scenario, an algorithmic model has been explained by incorporating a significant Recommendation System with an effective influence measurement process in order to generate a list of top most influenceable target users among all interacted network members for any specific social network user (main user).

At first the list of interacted network members (based on activities) has been updated with a less value by the operation of their social activities analysis based on contents classifcation with sentiment types and after that using the recommendation sytem of textual lexical analysis based on similarity and frequency, on which the interacted network members with most similar activities have been recommended with proper ranking by analyzing the recommendation of their post contents with respect to the post contents of the main user.

After that top most influenceable target network members (users) from that updated list of interacted network members have been identified for the main user based on any particular category and sentiment (example, influence of political category and positive sentiment) based on their social activities.

This algorithmic model can be cross validated by analyzing the results like similarly they have been justified in the previous research works [2, 3] done by us. Those testing procedures were very time consuming and also sometimes they are considered based on hypothesis [18, 19], these are some pioneering research works have already been successfully identified in these fileds.

## 9.0 Future Scope

Firstly in future the updated and large amount of dataset can be the most important direction for further result analysis which also can open few drawbacks or improvements for our algorithmic model with lot of eye opening justifications. This research also can be extended with more influence category classes with more sub domains and semantic analysis can be addition with sentiment analysis. Apart from that more natural language processing (nlp) analitical works such as personality analysis [3], emotion recognition etc based on social activities contents can be incorporated.

As we have considered only textual data, so considering images will be a vast field though some previously implemented methods like the *Optical Character Recognition* (OCR) on textual part recovery from images by Patel et al. [20] and the *Convolutional Neural Network* (CNN) for classifying patterns on images by Rakshit et al. [21]

Including these all above mentioned further scopes, there are also many uncovered fields which have already been mentioned in our previous research papers [2, 3], that can be great exposure for future.

**Footnote Links**

1. https://www.facebook.com/
2. https://twitter.com/